# Personalizing the meshed SPL/NAC Brain Atlas for patient-specific scientific computing using SynthMorph

Andy Huynh[1], Benjamin Zwick[1], Michael Halle[2], Adam Wittek[1], and Karol Miller[1]

[1] Intelligent Systems for Medicine Laboratory (ISML), The University of Western Australia, Perth, WA, Australia
`andy.huynh@research.uwa.edu.au`
[2] Surgical Planning Laboratory (SPL), Department of Radiology Brigham and Women's Hospital, and Harvard Medical School, Boston, MA, USA

**Abstract.** Developing personalized computational models of the human brain remains a challenge for patient-specific clinical applications and neuroscience research. Efficient and accurate biophysical simulations rely on high-quality personalized computational meshes derived from patient's segmented anatomical MRI scans. However, both automatic and manual segmentation are particularly challenging for tissues with limited visibility or low contrast. In this work, we present a new method to create personalized computational meshes of the brain, streamlining the development of computational brain models for clinical applications and neuroscience research. Our method uses SynthMorph, a state-of-the-art anatomy-aware, learning-based medical image registration approach, to morph a comprehensive hexahedral mesh of the open-source SPL/NAC Brain Atlas to patient-specific MRI scans. Each patient-specific mesh includes over 300 labeled anatomical structures, more than any existing manual or automatic methods. Our registration-based method takes approximately 20 minutes, significantly faster than current state-of-the-art mesh generation pipelines, which can take up to two hours. We evaluated several state-of-the-art medical image registration methods, including SynthMorph, to determine the most optimal registration method to morph our meshed anatomical brain atlas to patient MRI scans. Our results demonstrate that SynthMorph achieved high DICE similarity coefficients and low Hausdorff Distance metrics between anatomical structures, while maintaining high mesh element quality. These findings demonstrate that our registration-based method efficiently and accurately produces high-quality, comprehensive personalized brain meshes, representing an important step toward clinical translation.

## 1    Introduction

Computational models of the human brain are powerful tools that drive advances in both scientific discovery and clinical care. These models have enabled researchers and clinicians to better understand and predict brain behavior across diverse conditions.

However, when applied to different subjects, these models produce results that vary significantly [1], [2], [3], [4], highlighting the need for personalized brain models. Personalizing models remains a challenge due to the brain's inherent geometric complexity and anatomical variability between subjects. For computational brain models to achieve widespread clinical adoption, experts suggest that patient-specific models must be generated quickly and efficiently - specifically, producing quality meshes in under 40 minutes using standard computing hardware [5]. Currently, the specialized knowledge and effort required to develop these personalized models continues to limit their implementation in clinical practice and research.

Most previous work that has attempted to solve this problem have tried two approaches, segmentation-based model construction [3], [6], [7] and registration-based methods (also known as mesh or model morphing) [8], [9], [10], [11]. The segmentation-based approach is more involved and starts by applying a manual or automatic segmentation to the subjects MRI scan to extract anatomical labels. These labels are then used to create triangulated surfaces which are required for volumetric meshing. Once the volumetric mesh is created, the model is then developed by referencing the different anatomical structures. The morphing approach typically involves using an image registration method to compute the deformation field used to morph the brain mesh or model. Although this area of research has been extensively studied, current methods have several key limitations: (1) They work with limited anatomical structures, (2) use outdated registration techniques, and (3) are often locked to specific applications. This constrains both their utility in research and their adaptability to new clinical use cases.

We propose an approach that combines two methods used in previous work, segmentation-based model construction and registration-based methods, to streamline the development of patient-specific computational brain models. In our previous work [12], [13], we generated an anatomically labelled, hexahedral mesh from Open Anatomy Project's SPL/NAC Brain Atlas [14]. By leveraging this anatomical atlas, we constructed a comprehensive mesh with over 300 anatomical labels and ensured our mesh had high-quality elements, conforming boundaries at important interfaces and included the necessary anatomical labels required for efficient and accurate model development.

In this work, we present a method to morph the meshed anatomical brain atlas to a patient's MRI scan using medical image registration. We evaluated three state-of-the-art medical image registration methods to determine the registration method best suitable to morph our meshed anatomical brain atlas to patient MRI scans. We assessed registration accuracy using DICE similarity coefficient and Hausdorff Distance. To assess the element quality after morphing the meshed anatomical brain atlas, we used established element quality metrics: scaled Jacobian, aspect ratio and skew. We demonstrate that our registration-based method efficiently and accurately produces high-quality, comprehensive personalized brain meshes, representing an important step toward clinical translation and enabling research.

## 2    Method

### 2.1    Datasets

#### 2.1.1 Open Anatomy Project's SPL/NAC Brain Atlas

The SPL/NAC digital brain atlas was developed through collaboration between Brigham and Women's Hospital and Massachusetts General Hospital (Fig. 1). The atlas is based on MRI scans of a healthy 42-year-old male volunteer, acquired at the Martinos Center for Biomedical Imaging using a Siemens 3T scanner with a multi-array head coil. The imaging protocol consisted of T1-weighted (MPRAGE) and T2-weighted sequences, both acquired at 0.75 mm isotropic resolution. The atlas features comprehensive anatomical labeling of over 300 structures, generated through a combination of Freesurfer's automatic parcellation and extensive manual segmentation. These labels are provided in 1 mm isotropic resolution volumes and organized according to the Radlex ontology framework. The complete atlas dataset, including volumetric MRI scans and corresponding anatomical labels, is freely accessible through the Open Anatomy Project platform [14], [15].

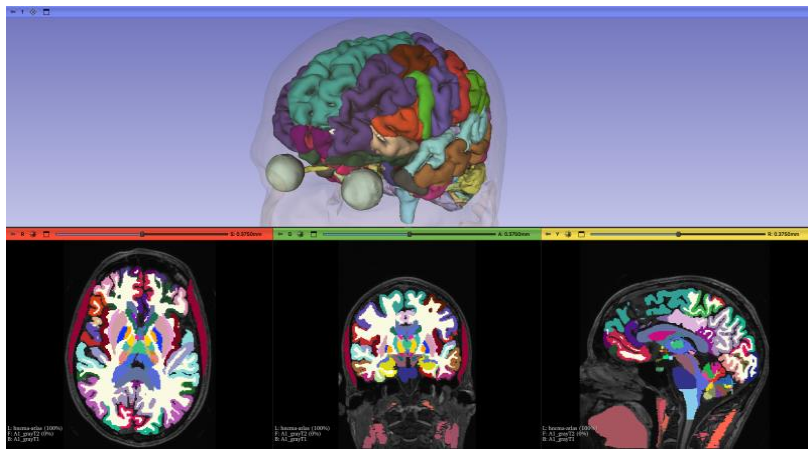

**Fig. 1.** Open Anatomy Project's SPL/NAC Brain Atlas [14], [15]. Top view is the 3D model rendering of the anatomical label map and the bottom view is the label map overlaid onto the subject's MRI scan.

#### 2.1.2 Test subject MRI dataset

We used the publicly available image dataset [16] that consists of multimodal MRI scans acquired using a Philips Medical Systems 3T scanner. The imaging protocol included T1-weighted imaging without contrast (voxel size: $0.469 \times 0.469 \times 0.939$ mm, matrix size: $512 \times 512 \times 200$), T2-weighted spin-echo imaging (voxel size: $0.429 \times 0.429 \times 3$ mm, matrix size: $560 \times 560 \times 55$), and diffusion-weighted imaging with 32 non-collinear gradient directions (voxel size: $0.469 \times 0.469 \times 0.939$ mm, matrix size: $512 \times 512 \times 200$).

This dataset was selected due to its association with an experiment involving unilateral median nerve stimulation which will be applied in our subsequent research.

### 2.2 Construction of meshed anatomical brain atlas

We developed an anatomically detailed hexahedral mesh using Open Anatomy Project's SPL/NAC Brain Atlas, referred to as the meshed anatomical brain atlas [12], [13] (Fig. 2). Our construction process consisted of three primary stages:

**Pre-processing**. We created an additional label map derived from the original atlas that incorporates structures essential for finite element modeling. To create this, we developed an extension in open-source software 3D Slicer called SlicerAtlasEditor [17], for manipulation of the SPL/NAC Brain Atlas. This was used to identify and label key anatomical components, such as grey matter, white matter, and ventricular system. We used SynthStrip [18], a skull-stripping tool, to identify and label the scalp, skull and cerebrospinal fluid (CSF). This additional label map is referred to as the 'Material label map'.

**Mesh generation**. We created smooth outer boundary triangulated surfaces by extracting the scalp, skull, and CSF interfaces using a marching cube algorithm. These surfaces then guided the generation of conforming hexahedral meshes using an overlay-grid procedure. Mesh element quality assessment confirmed that both meshes exceeded recommended standards [19], [20], with over 97% of elements showing excellent scaled Jacobian, aspect ratio, and skew metrics.

**Mesh labeling.** We used an octree-based spatial search method, implemented in the VTK library [21], to accurately transfer the SPL/NAC Brain Atlas and the custom 'Material label map' to the hexahedral mesh. This ensured precise anatomical structure labelling, while maintaining mesh quality for finite element analysis.

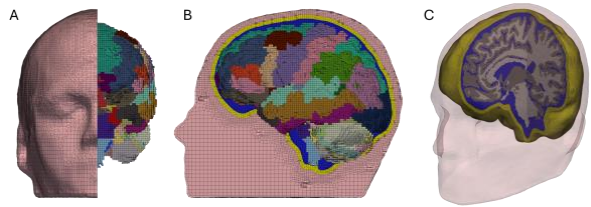

**Fig. 2.** The meshed anatomical brain atlas. A) Fontal view of the meshed atlas cross-section with the SPL/NAC brain atlas labels. B) Side view of the meshed atlas cross-section with the SPL/NAC brain atlas labels. C) The meshed atlas cross-section with the Material labels.

### 2.3 Morphing of meshed anatomical brain atlas

We developed a framework to morph our meshed anatomical brain atlas [12], [13], [22] to match an individual's brain MRI scan (see Fig. 3):

**(a) Medical image registration.** Medical image registration is used to acquire the transform that maps the atlas MRI space to the target MRI space. Three different registration methods were compared to identify the optimal method. We compared three types of image registration methods: a global affine registration method, a local B-Spline registration method (both implemented in SimpleElastix/SimpleITK [23], [24]) and a machine learning method called SynthMorph [25], [26].

**(b) Image registration evaluation.** To evaluate the accuracy of the transforms computed from each method, we conducted a similarity assessment. We quantified the anatomical accuracy of the registration using DICE Similarity Coefficient (DICE) and Hausdorff Distance (HD) metrics between corresponding structures in the atlas and subject. We computed these metrics for a variety of anatomical structures that were segmented using FreeSurfer [27]. We used FreeSurfer to segment the atlas MRI scan and the target MRI scan. We then used the transform outputted from each of the image registration method to morph the segments of the MRI scan that was used to create the atlas. We then compared the morphed segmented anatomical structures of the atlas MRI scan and the 'ground truth' segmented anatomical structures of the target MRI scan.

**(c) Mesh morphing.** Each method generated a transform, mapping the atlas to the target's MRI space. The transforms were applied to the meshed anatomical brain atlas, resulting in three morphed meshes (affine, B-Spline and SynthMorph).

**(d) Mesh element quality evaluation.** We assessed the quality of the morphed mesh elements using scaled Jacobian, skewness, and aspect ratio measurements to ensure computational reliability. Through this comprehensive evaluation, we identified the registration method that provided the optimal balance between anatomical accuracy and mesh quality preservation. We compared the mesh element quality of the meshed anatomical brain atlas (reference) and the morphed meshes (affine, B-Spline and SynthMorph). The reported metrics included the scaled Jacobian, aspect ratio and skew metrics. These metrics were computed using VTK library's implementation of the Verdict library [19]. It should be noted that there are no universally accepted quality standards for the mesh as there are multiple ways to judge element quality. We follow recommendations from both the Verdict library [19] and King Yang [20].

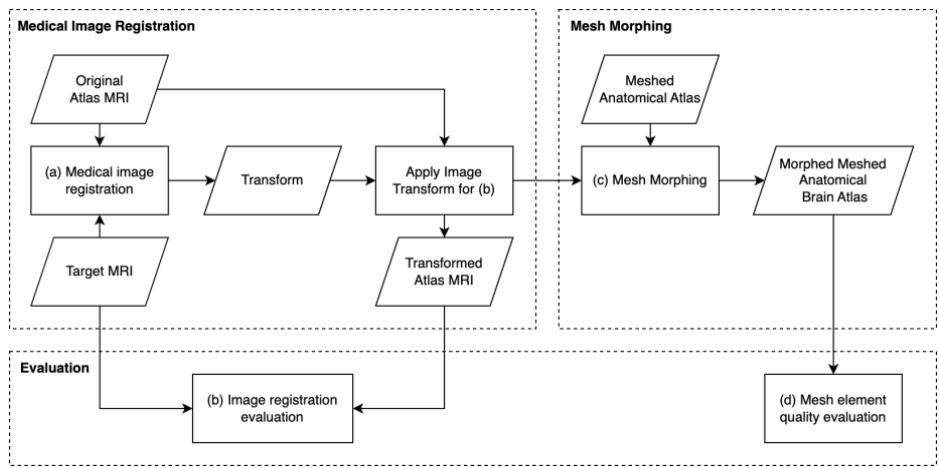

**Fig. 3.** Flowchart describing the morphing of the meshed anatomical brain atlas.

## 3 Results and Discussion

### 3.1 Assessment of accuracy of image registration

Our registration results are presented in Fig. 5, showing the transform overlayed on the resulting transformed atlas MRI scan. Our numerical results demonstrate the relative performance of three registration methods, as shown in Table 1. SynthMorph achieved superior performance across all evaluation metrics, with a DICE coefficient of 0.760, substantially outperforming both Affine (0.556) and B-Spline (0.516) approaches. The improvement is particularly notable in terms of the Hausdorff Distance (HD), where SynthMorph obtained an HD of 5.248 mm, representing approximately a 37% registration accuracy improvement compared to Affine registration (8.351 mm) and a 43% reduction compared to B-Spline (9.180 mm). Similarly, for the 95-percentile Hausdorff Distance (HD95), SynthMorph maintained its advantage with HD95 of 2.245 mm, showing markedly better performance than both Affine (4.200 mm) and B-Spline (4.532 mm) methods. These results suggest that SynthMorph provides more accurate and robust registration compared to traditional approaches.

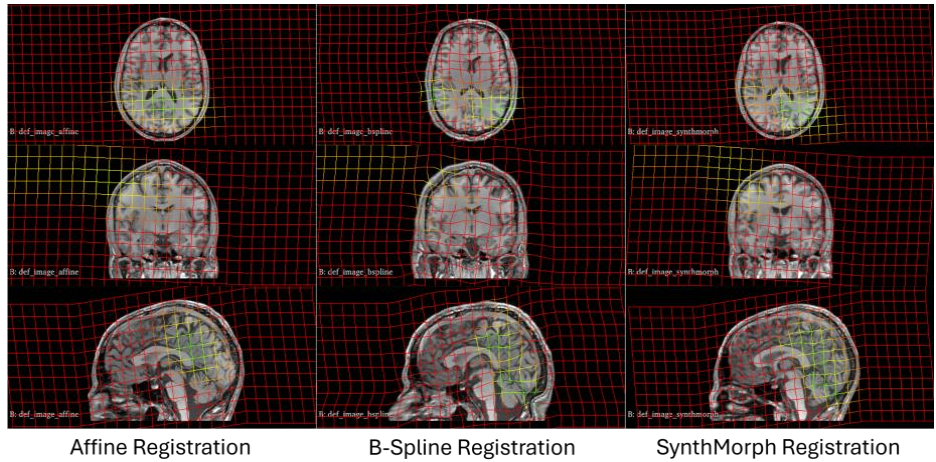

**Fig. 4.** Registration results with deformation field (transform) overlay on transformed atlas MRI scan. Top row: axial view, middle row: sagittal view and bottom row: coronal view.

**Table 1.** Registration evaluation results for different methods, including the DICE coefficient, Hausdorff Distance and 95-percentile Hausdorff Distance. Best results highlighted in bold.

| Methods | DICE | HD (mm) | HD95 (mm) |
| --- | --- | --- | --- |
| Affine | 0.556 | 8.351 | 4.200 |
| B-Spline | 0.516 | 9.180 | 4.532 |
| SynthMorph | **0.760** | **5.248** | **2.245** |

### 3.2 Assessment of quality of meshed anatomical atlas

The mesh quality metrics reveal distinctive patterns across different registration methods. The reference (unmorphed) mesh demonstrates high quality baseline characteristics, with 97.46% of elements having scaled Jacobian values (1 is the best possible value) above 0.5 and an average of 0.911. The aspect ratio (1 is the best possible value) shows 87.99% of elements below 3, with an average of 1.569. The skew metric (lower the better) remains predominantly low with 99% of elements below 0.5 and an average of 0.095.

SynthMorph preserves 96.95% of elements with scaled Jacobian above 0.5 and a decrease in average value to 0.864. The aspect ratio distribution is maintained with 88.17% of elements below 3, and a higher average of 1.726. The method's skew metrics reduces to 95.13% elements below 0.5 and an increase average value to 0.248. Overall, the element quality remains reasonable with slight reduction compared to the reference mesh.

Comparing registration approaches, the affine transformation maintains mesh quality closest to the reference, with nearly identical scaled Jacobian statistics and minimal changes in aspect ratio and skew distribution. However, it results in poor anatomical alignment. The B-Spline method shows similar element quality to the SynthMorph method, most likely due to both being local image registration methods. This suggests that while SynthMorph and B-Spline methods result in similar element quality, SynthMorph is the better choice due to appreciably more accurate anatomical alignment.

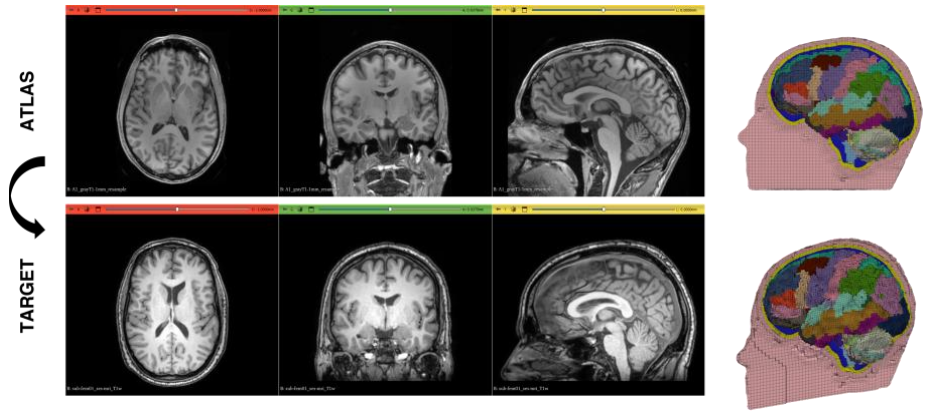

**Fig. 5.** Morphed meshed anatomical atlas. Top row: atlas MRI scan with OAP's SPL/NAC brain atlas label map overlaid (left), and the meshed anatomical atlas (right). Bottom row: target brain MRI scan (left), and the morphed meshed anatomical atlas using SynthMorph (right).

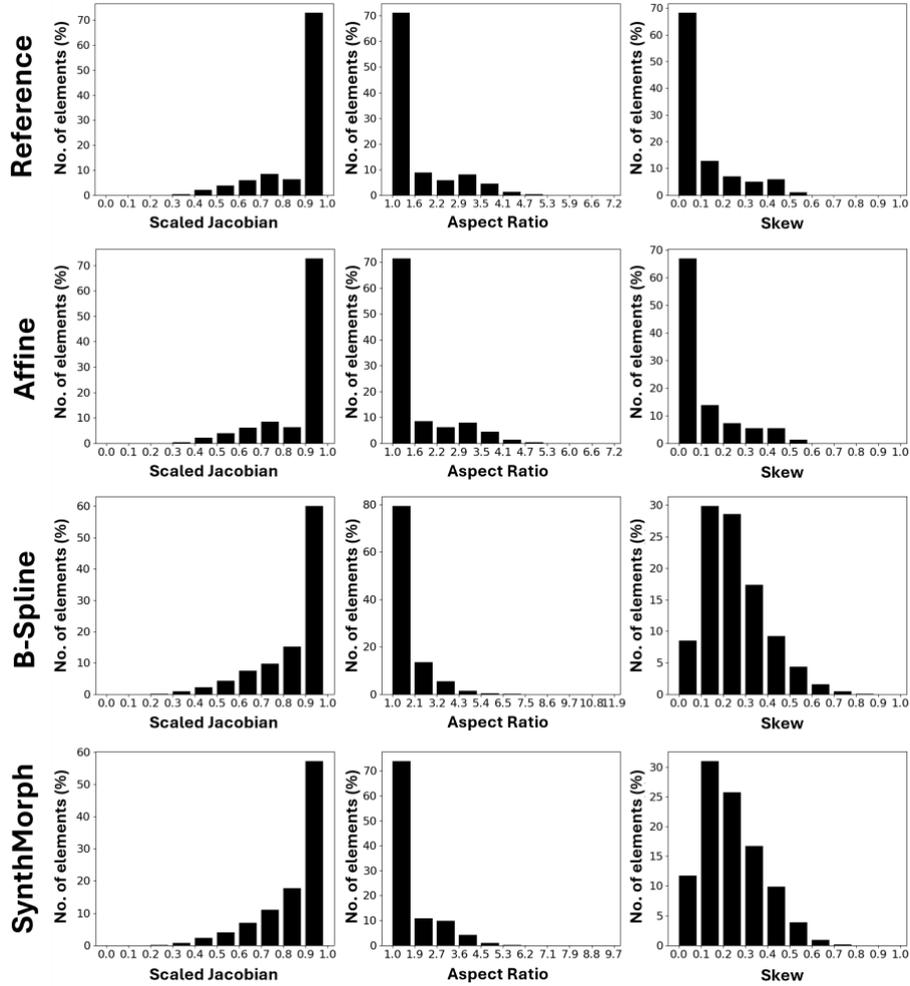

**Fig. 6.** Distribution of mesh element quality metrics of the meshed anatomical atlas (reference) and the morphed meshed anatomical atlas (using affine, b-spline and SynthMorph).

## 4 Conclusions

We developed a novel agnostic image registration framework for morphing the meshed anatomical brain atlases that can be used with any image registration method. We evaluated three image registration methods to find the best suited method considering anatomical accuracy and preservation of mesh element quality. SynthMorph provided better anatomical accuracy with the cost of lesser element quality. Affine registration had negligible effects on element quality, but resulted in significantly worse anatomical accuracy. Lastly, B-Spline registration performs poorly in both, which may be because it requires manual calibration of the algorithm for it to perform well.